\documentclass[manuscript,screen]{acmart}
\AtBeginDocument{%
  \providecommand\BibTeX{{%
    \normalfont B\kern-0.5em{\scshape i\kern-0.25em b}\kern-0.8em\TeX}}}

\setcopyright{acmlicensed}
\copyrightyear{2024}
\acmYear{2024}
\setcopyright{acmlicensed}\acmConference[CHI '24]{The Methods for Family-Centered Design Workshop at the CHI Conference on Human Factors in Computing Systems}{May, 2024}{Honolulu, HI, USA}
\acmBooktitle{The Methods for Family-Centered Design Workshop at the CHI Conference on Human Factors in Computing Systems (CHI '24), May, 2024, Honolulu, HI, USA}

\newcommand{\out}[1]{{#1}}

\newcommand{\summary}[1]{\out{{\textcolor{blue}{\textbf{ }}}}}

\begin{document}

\title{Generating A Crowdsourced Conversation Dataset to Combat Cybergrooming}

\author{Xinyi Zhang}
\email{xinyizhang@vt.edu}
\affiliation{%
  \institution{Virginia Tech}
  \city{Blacksburg}
  \state{Virginia}
  \country{USA}
}

\author{Pamela J. Wisniewski}
\email{pamela.wisniewski@vanderbilt.edu}
\affiliation{%
  \institution{Vanderbilt University}
  \city{Nashville}
  \state{Tennessee}
  \country{USA}
}

\author{Jin-Hee Cho}
\email{jicho@vt.edu}
\affiliation{%
  \institution{Virginia Tech}
  \city{Blacksburg}
  \state{Virginia}
  \country{USA}
}

\author{Lifu Huang}
\email{lifuh@vt.edu}
\affiliation{%
  \institution{Virginia Tech}
  \city{Blacksburg}
  \state{Virginia}
  \country{USA}
}

\author{Sang Won Lee}
\email{sangwonlee@vt.edu}
\affiliation{%
  \institution{Virginia Tech}
  \city{Blacksburg}
  \state{Virginia}
  \country{USA}
}

\renewcommand{\shortauthors}{Zhang, et al.}

\begin{abstract}
\section{Abstract}

Cybergrooming emerges as a growing threat to adolescent safety and mental health. One way to combat cybergrooming is to leverage predictive artificial intelligence (AI) to detect predatory behaviors in social media. However, these methods can encounter challenges like false positives and negative implications such as privacy concerns. Another complementary strategy involves using generative artificial intelligence to empower adolescents by educating them about predatory behaviors. To this end, we envision developing state-of-the-art conversational agents to simulate the conversations between adolescents and predators for educational purposes. Yet, one key challenge is the lack of a dataset to train such conversational agents. In this position paper, we present our motivation for empowering adolescents to cope with cybergrooming. We propose to develop large-scale, authentic datasets through an online survey targeting adolescents and parents. We discuss some initial background behind our motivation and proposed design of the survey, such as situating the participants in artificial cybergrooming scenarios, then allowing participants to respond to the survey to obtain their authentic responses. We also present several open questions related to our proposed approach and hope to discuss them with the workshop attendees. 
\end{abstract}

\begin{CCSXML}
<ccs2012>
 <concept>
  <concept_id>00000000.0000000.0000000</concept_id>
  <concept_desc>Do Not Use This Code, Generate the Correct Terms for Your Paper</concept_desc>
  <concept_significance>500</concept_significance>
 </concept>
 <concept>
  <concept_id>00000000.00000000.00000000</concept_id>
  <concept_desc>Do Not Use This Code, Generate the Correct Terms for Your Paper</concept_desc>
  <concept_significance>300</concept_significance>
 </concept>
 <concept>
  <concept_id>00000000.00000000.00000000</concept_id>
  <concept_desc>Do Not Use This Code, Generate the Correct Terms for Your Paper</concept_desc>
  <concept_significance>100</concept_significance>
 </concept>
 <concept>
  <concept_id>00000000.00000000.00000000</concept_id>
  <concept_desc>Do Not Use This Code, Generate the Correct Terms for Your Paper</concept_desc>
  <concept_significance>100</concept_significance>
 </concept>
</ccs2012>
\end{CCSXML}


\keywords{Do, Not, Us, This, Code, Put, the, Correct, Terms, for,
  Your, Paper}


\received{20 February 2007}
\received[revised]{12 March 2009}
\received[accepted]{5 June 2009}


\begin{abstract}

Cybergrooming emerges as a growing threat to adolescent safety and mental health. One way to combat cybergrooming is to leverage predictive artificial intelligence (AI) to detect predatory behaviors in social media. However, these methods can encounter challenges like false positives and negative implications such as privacy concerns. Another complementary strategy involves using generative artificial intelligence to empower adolescents by educating them about predatory behaviors. To this end, we envision developing state-of-the-art conversational agents to simulate the conversations between adolescents and predators for educational purposes. Yet, one key challenge is the lack of a dataset to train such conversational agents. In this position paper, we present our motivation for empowering adolescents to cope with cybergrooming. We propose to develop large-scale, authentic datasets through an online survey targeting adolescents and parents. We discuss some initial background behind our motivation and proposed design of the survey, such as situating the participants in artificial cybergrooming scenarios, then allowing participants to respond to the survey to obtain their authentic responses. We also present several open questions related to our proposed approach and hope to discuss them with the workshop attendees. 
\end{abstract}

\maketitle

\section{Introduction and Background}

Cybergrooming is a long-term online activity in which a predator, usually an adult, befriends adolescents, lures them with some benefits, and eventually solicits them for sexual encounters both online and offline~\cite{mladenovic2021cyber}. As cybergrooming becomes more prevalent online, it has started to make a significant impact on adolescent’s safety and health. 
Research has shown that this predatory behavior negatively impacts youth mental health and safety, leading to long-term side effects such as anxiety, depression, post-traumatic stress, and suicide~\cite{INHOPE}. 

Researchers have explored different ways to combat cybergrooming. 
For example, recently, with the rise of conversation agents powered by predictive artificial intelligence (AI), researchers have also explored various strategies to use algorithms to detect potential risks during the conversation between predators and victims~\cite{bours2019detection}. 
However, these approaches primarily focus on detecting and mitigating risks and cannot be entirely accurate due to the complex and ambiguous nature of such conversations, and may yield misses or false positives. 
In addition, such detection mechanisms come with privacy concerns for individuals, as the algorithm needs to constantly monitor online communication.
A complementary approach is to enhance coping mechanisms and empower youth to recognize the risks of cybergrooming by educating adolescents to be resilient. 

The rapid advancement of generative AI provides new opportunities to empower youth using conversational agents. 
For instance, conversational agents can provide a simulation environment for youth to experience different kinds of cybergrooming without exposing them to actual risks in the real world~\cite{wang2023authentic}. 
However, one challenge to developing such conversation agents is to access datasets that provide adequate instances of authentic cybergrooming incidents. 
The existing dataset that researchers~\cite{gupta2012characterizing, gunawan2016detecting, mladenovic2021cyber, guo2023text} often rely on is the Perverted Justice (PJ) Dataset, a public conversation corpus between predators and victims disguised by volunteers. Yet, this dataset has some limitations. One limitation is that adolescents in the real world may behave differently from the ones who pose as adolescents whose goal is to lure predators. In addition, the PJ dataset was generated between 2004 and 2016, which makes it outdated to reflect how adolescents communicate in modern social media. Another limitation is that the PJ dataset only has vulnerable behaviors and without any resilient behaviors. Therefore, we do not have a dataset on how predators respond to circumvent the resilient responses from adolescents. These limitations make it challenging to create an authentic and comprehensive conversation agent that can demonstrate both adolescents' resilient and vulnerable behaviors and how predators respond to such behaviors. 

To that end, we propose leveraging both parents and adolescents in generating data for a cybergrooming corpus. Parents naturally have a strong instinct to protect their children from harm which could translate into proactive behaviors. Therefore, we believe that parents can be harnessed to demonstrate resilient behaviors. Additionally, it provides an opportunity to understand the discrepancy between the two groups in how they perceive vulnerable and resilient behaviors differently.
In this position paper, we share how we plan to involve both parents and adolescents to generate a dataset that can be used to create large-scale, authentic cybergrooming conversation datasets. Such datasets will lay a solid foundation for future research in developing educational conversation agents or designing other educational interfaces to empower youth to combat cybergrooming.

\begin{figure}
    \centering
    \includegraphics[width=0.35\textwidth]{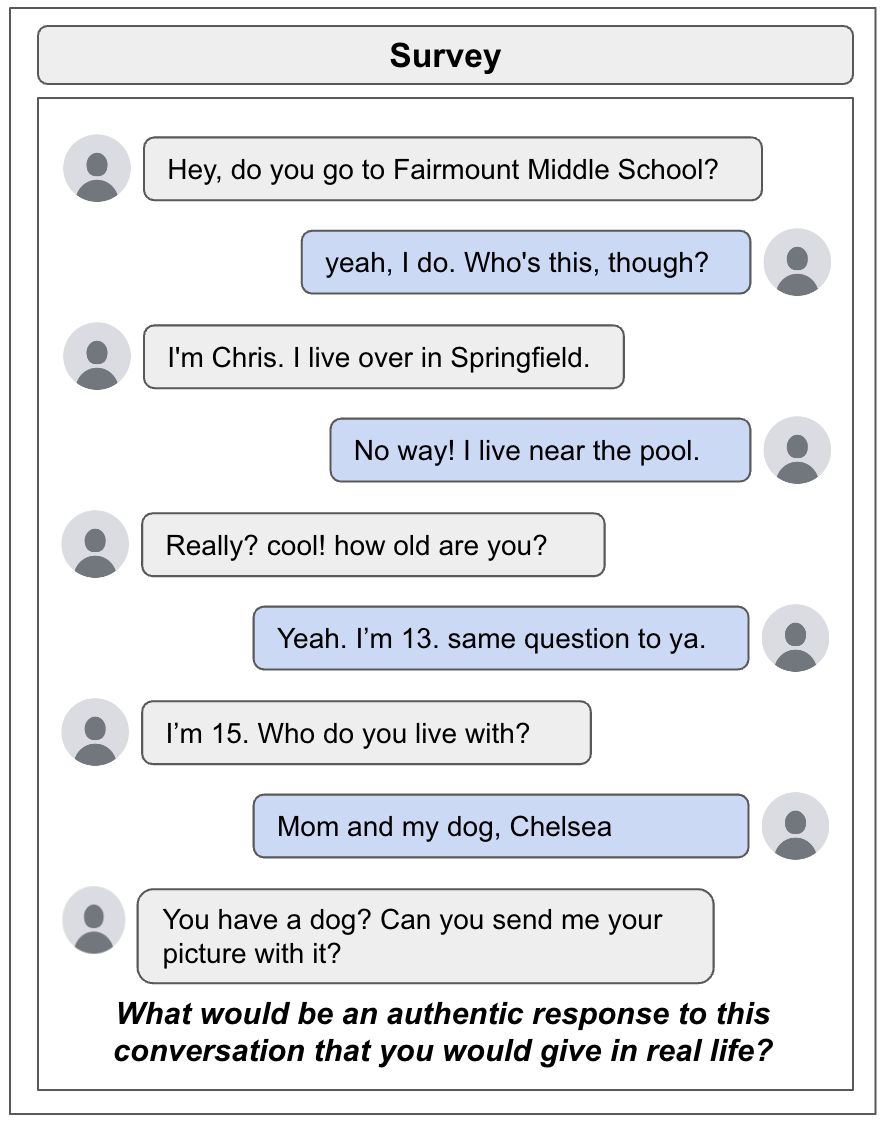}
    \caption{An example conversation scenario for the proposed survey}
    \label{fig:enter-label}
\end{figure}

\section{Proposed Work: Parents/Adolescents sourcing}
In our research, we propose crowdsourcing conversation responses from parents and adolescents in simulated cybergrooming scenarios through an online survey. We believe incorporating parents' and adolescents' perspectives is important for a more comprehensive understanding of cybergrooming. Adolescents will provide authentic responses, while parents' responses can help us identify effective risk-coping responses with stronger motivation to protect their children. These data can be used to train generative NLP models capable of providing conversations close to authentic cases. Parents and adolescents can further provide complementary responses in many cases. For example, given the same cybergrooming conversation scenario, parents may consider it risky, while adolescents may treat them as safe. Such discrepancy between parents and adolescents provides opportunities to carry out more targeted education. Lastly, through the process of generating data, parents and adolescents will have more awareness of how predators communicate with adolescents. 

In our survey, we plan to present participants --- both parents and adolescents --- conversations between victims and predators of cybergrooming. Then we will ask participants to respond to the conversations in two different categories: 1) vulnerable behaviors that can elevate the risks of being groomed or vulnerable,  and 2) resilient behaviors that can protect themselves from the risk. Additional follow-up questions can be used to rate how much they perceive the risks from the given scenarios. Such survey example is shown in Fig.~\ref{fig:enter-label}).
The team has experience involving adolescents for data collection purposes~\cite{razi2023sliding, razi2022instagram}, as well as co-design~\cite{agha2023strike, agha2023co} and is well aware of potential negative implications that should be considered when involving adolescents. 
For example, we do plan to recruit parents and adolescents separately rather than jointly, as it can impact the authenticity of their responses.

The collected data will be analyzed to identify the perceived differences between the two groups in various components of cybergrooming: perceived risks, resilient behaviors, and vulnerable behaviors. We will use qualitative research methods. In addition, we plan to understand the linguistic characteristics of four different types of data: parents' vulnerable behaviors, parents' resilient behaviors, adolescents' vulnerable behaviors, and adolescents' resilient behaviors Our analysis will provide insights into developing coping skills for cybergrooming and use them to develop conversational agents.
The collected dataset will be used to develop conversational agents. 

\section{Open questions to discuss at the workshop} 
In the workshop, we would like to discuss the following questions with other workshop attendees:

\begin{itemize}
    \item What is the effective recruitment strategy for involving adolescents and parents in generating the dataset for cybergrooming?
    \item Compared to traditional crowdsourcing, how should we approach our participants (parents and adolescents) differently, considering the research context of cybergrooming? 
    \item What ethical considerations should we take into account when collecting such a dataset?
    \item How should we balance the authenticity of conversation scenarios with the protection of adolescents from exposure to sensitive content? 
\end{itemize}

\bibliographystyle{ACM-Reference-Format}
\bibliography{cybergrooming}

% \appendix

\end{document}